# Thermal properties study of silicon nanostructures by photoacoustic techniques

K. Dubyk[1], T. Nychyporuk[2], V. Lysenko[3,4,] K. Termentzidis[5], G. Castanet[6], F. Lemoine[6], D. Lacroix[6], M. Isaiev[6]

[1] *Taras Shevchenko National University of Kyiv, 64/13, Volodymyrska str., 01601 Kyiv, Ukraine*
[2] *Université de Lyon; Institut des Nanotechnologies de Lyon INL-UMR5270, CNRS, INSA de Lyon, Villeurbanne F-69621, France*
[3] *Nanotechnology Institute of Lyon (INL), UMR CNRS 5270, University of Lyon, 69361 Lyon, France;*
[4] *Light Matter Institute, UMR-5306, Claude Bernard University of Lyon/CNRS, Université de Lyon 69622 Villeurbanne cedex, France*
[5] *Université de Lyon, CNRS, CETHIL, Villeurbane, F-69000, France*
[6] *Université de Lorraine, CNRS, LEMTA, Nancy, F-54000, France*

The photoacoustic method with piezoelectric detection for the simultaneous evaluation of the thermophysical properties is proposed. The approach is based on the settling of an additional heat sink for redistribution of heat fluxes deposited on the sample surface. Firstly, the approach was tested on the porous silicon with well-defined morphology and well-studied properties. Then, heat capacity and thermal conductivity of silicon nanowires arrays have been investigated by recovering the experimental data through numerical simulations. The decrease of heat capacity and effective thermal conductivity of the samples upon increasing thickness and porosity of the sample is observed. Such behavior could be caused by the increase of the structure heterogeneity. In particular, this can be related to larger disorder (increased density of broken nanowires and larger porosity) that appears during the etching process of the thick layers.

## 1. Introduction

Silicon-based materials are "sine qua non" in micro/nano/opto-electronics[1–5]. The continuous size reduction of such devices and the growing aspect of their efficient integration naturally lead to improve thermal management. Therefore, understanding heat transfer phenomena at the nano and micro scale is crucial for performance stability and reliability of nano/micro devices.

Numerous in-contact and out-of-contact experimental methods are used in the literature for thermal transport properties studies. With in-contact configurations, the sample is directly heated with a source which can be either a microprobe as in scanning thermal microscopy [6,7], a thin-film electrical resistor deposited onto the surface of the sample as in 3ω-method [8,9], a bolometer [10] as in pulse heating regime [11], or an external resistor in the case of hot disk [12]. These techniques can be applied for measuring thermal properties of different structures such as bulk, thick and thin films. The main disadvantage of these techniques is the presence of thermal contact resistance between the heat source and the sample. Therefore, use of such techniques for the study of samples with rough surfaces, like nanowires arrays, is very challenging.

Non-contact methods are based on the remote excitation of the thermal perturbation in the material. For instance, in photothermal methods, the thermal perturbation is induced by light irradiation. In this frame there is no contact resistance, and this is the significant advantage of such methods, which do not require special pre-treatment of the studied sample. On the other hand, it is difficult to evaluate the amount of energy absorbed in the sample. Consequently, several issues can be expected when material heat capacity measurements are required.

In practice, methods with transient excitation of thermal perturbations allow the determination thermal diffusivity ($D$). Which relates to thermal conductivity ($\kappa_s$) with the following equation:

$$D = \kappa_s / c\rho,$$

where $c$ is the specific heat and $\rho$ is the density of the material. Therefore, to evaluate the thermal conductivity, both density and specific heat should be known, which is challenging for nanostructured materials.

Often, the density and the specific heat for nanostructures are considered to be the same as for the bulk state[13,14]. However, this assumption is crude, since changes in phonon dispersion and influence of phonon surface modes can affect the specific heat [15]. Generally, features of the structure determine the application of the assumptions[16], while a universal approach applicable for all case does not exist. Consequently, the development of new methods for the thermophysical properties study of nanostructured materials remains an important issue.

In this paper, the study of thermal conductivity of nanostructured materials with the use of a photoacoustic technique in piezoelectric configuration is presented [17–22]. An approach of simultaneous evaluation of thermal conductivity and heat capacity of material is proposed [23–25]. Then, materials as: porous silicon samples (PS) and silicon nanowires arrays (SiNWs), fabricated by electrochemical and metal-assisted chemical etching of a bulk silicon, are studied with the developed photoacoustic technique.

2. **Experimental details**
2.1. **Sample fabrication**

*Porous silicon.* The samples were fabricated by electrochemical etching of p+-type monocrystalline Si wafer doped with boron with a resistivity about 0.01-0.02 Ω·cm. The thickness of silicon wafer was 500 µm. The etching solution consists the hydrofluoric acid and ethanol (HF (49%):$C_2H_5OH$=1:1). For obtaining porous silicon with porosity $P$=45%, 55%, 65% different anodic current densities $j$=50 mA/cm$^2$, 100 mA/cm$^2$, 200 mA/cm$^2$, respectively, were applied. The etching time was set such to obtain the same thickness of all fabricated porous layers, which is measured to be equal to 50 µm. The morphologies of resulting porous silicon layers were observed via scanning electron microscopy (SEM). SEM image of the prepared porous silicon sample with porosity 45% is shown in Fig.1.

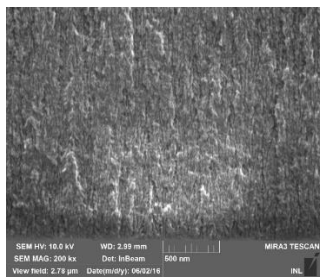

Fig.1. Cross sectional SEM image of porous silicon layer with porosity $P$=45% fabricated by electrochemical etching.

*Silicon nanowires arrays.* Samples of silicon nanowires arrays were fabricated by the method of metal-assisted chemical etching (MACE) of a monocrystalline p-type silicon substrate doped with boron. The initial thickness of the silicon wafer was 300 μm and resistance of initial wafer was 0,04-0,05 Ω·cm. Schematic illustration of the sucessive steps of the metal-assisted chemical etching of silicon nanowires arrays is shown in Fig. 2a.

The wafers were cleaned in an aqua solution of 5% hydrofluoric acid (HF) to remove oxide layer from substrates. After cleaning, substrates were immersed in a solution of 5 M hydrofluoric acid and 0.02 M silver nitrate. As a result, Ag nanoparticles were deposited onto the wafer surface. Next step is the formation of Si nanowires. For this purpose, the samples were immersed in an etching solution of 5M hydrofluoric acid and 30% hydrogen peroxide. Then, to dissolve the Ag dendrites the samples were immersed into 65% $HNO_3$. After that, it was rinsed with deionized water. Three series of the sample were fabricated with different etching time (35 min, 60 min and 90 min), which leads to formation of SiNWs with different thicknesses (20 μm, 35 μm and 50 μm, respectively). The porosity of fabricated SiNWs arrays are equal to (55±5) %. The typical SEM image of a fabricated sample with the thickness of SiNWs arrays equals to 50 μm is shown in Fig. 2b.

Porosity for all samples was firstly estimated with gravimetric method, and then more precise control with the use of SEM images.

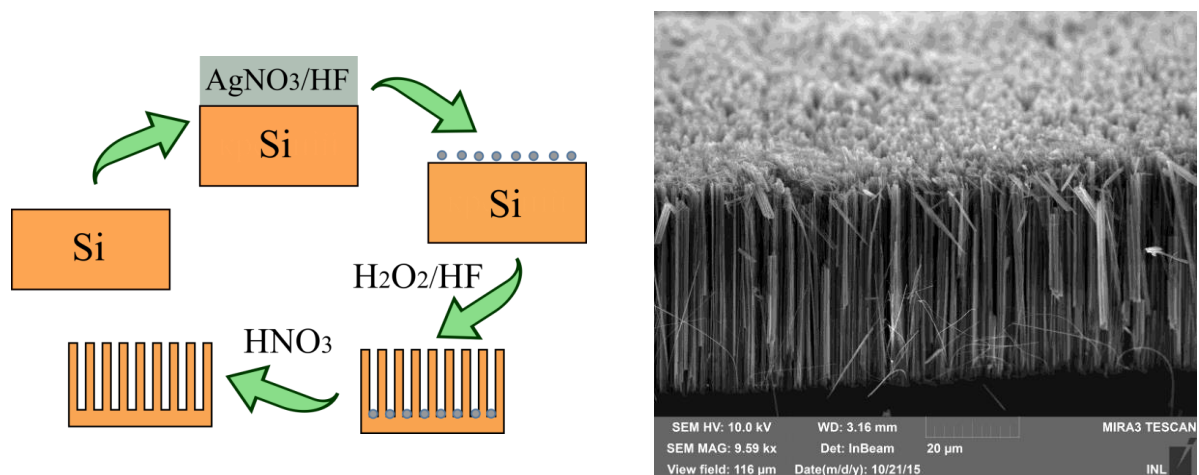

Fig.2 The sketch of MACE process stages (a); SEM image of fabricated sample (b).

### 2.2. Photoacoustic measurements

*Setup details*. As it was mentioned above, we used the photoacoustic technique with piezoelectric recording. Piezoelectric detection principle has several advantages while compared to the more classical gas-microphone one, such as larger working frequency range. Additionally, it gives the possibility to study photothermal response directly from the sample. Schematic configuration of the photoacoustic measurement unit with a piezoelectric detection system is shown in Fig. 3. The experiments were carried out in presence and absence of a covering liquid layer (water) with thickness equals to 1 mm. The samples were irradiated by modulated laser radiation with a wavelength equals to 532 nm and the electric power of 2 W. The square-wave modulation of the laser was performed by modulation of its power source with the Tektronix AFG1022 generator.

The parallel light beam was formed by the optical collimator system to form a spot on the sample surface with the size equal to 3 mm. The absorption of the electromagnetic radiation induced the heating of the sample. As a result, thermoelastic stresses were arising in the considered multilayered system. These stresses were detected by a piezoelectric transducer (PIC-151). Specific buffer layer with low thermal conductivity and thermal expansion coefficient was chosen for localization of thermoelastic sources in the studied sample and thus avoiding heat energy penetration into the piezoelectric transducer. This buffer was fabricated from glass ceramic layer (thickness is equal to 1 mm). The length and width of both the sample buffer layer and the PZT was 5 mm×5 mm. Photoacoustic responses were recorded in the frequency range from 10 Hz to 20 kHz. The typical amplitude-frequency characteristics are presented in result section (see Fig. 4).

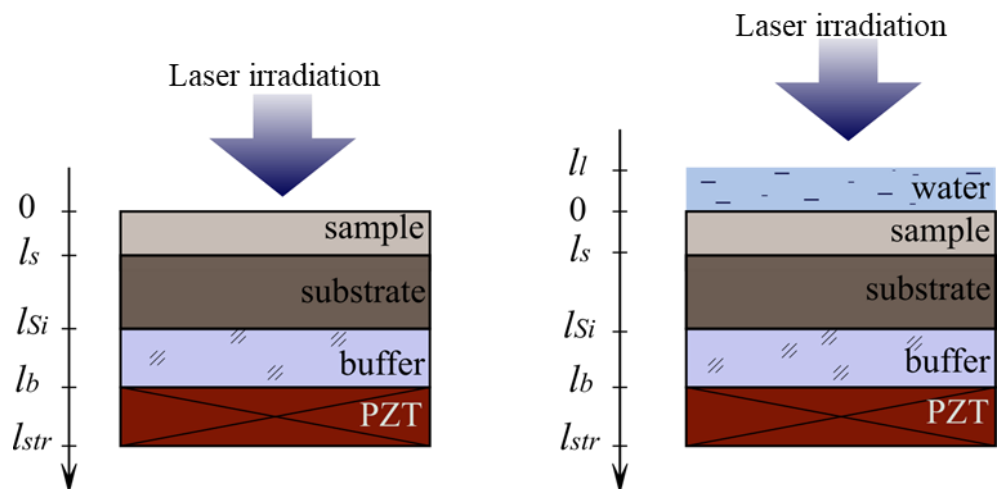

Fig. 3 Schematic sketch-view of a multilayered system in the case of absence (a) and presence (b) of the covering liquid layer.

*Analytic model*. Let us consider the photoacoustic response formation in multilayered structure shown in Fig. 3. The analysis of experimental amplitude-frequency characteristics was carried out in frames of quasi-stationary deformations assumption. The spatial distribution of thermoelastic stresses which arise in the studied structure can be represented as follows [26]

$$\sigma(z) = \frac{E(z)}{1-\nu(z)} \int_0^{l_{str}} G(z)\, \sigma_s(z),$$

here $E(z)$, $\nu(z)$ are the spatial distributions of Young's modules and Poisson's coefficient, $G(z)$ is the Green's function, $\sigma_s(z)$ is the thermoelastic stress source term. In the considered case, the source can be expressed as a thermoelastic force that depends on the temperature field in the layers:

$$\sigma_s(z) = \frac{E(z)}{1-\nu(z)} \alpha_T(z)\theta(z,\omega),$$

here $\alpha_T(z)$ is the thermal expansion coefficient of the material, $\theta(z,\omega)$ is the spatial distribution of variable component of temperature. The spatial distribution of the temperature in multilayered structure can be evaluated by solving heat conduction equation with the appropriate boundary conditions[27]:

$$c\rho \frac{\partial \theta}{\partial t} = \frac{\partial}{\partial z}\left(\kappa(z)\frac{\partial \theta}{\partial z}\right) + I\alpha e^{-\alpha z}$$

here $I$ is the intensity of the absorbed radiation, $\alpha$ is the optical absorption coefficient.

It should be noted that the effect of electron-hole recombination was neglected in heat conduction equation, since the lifetime of carriers in porous silicon and silicon nanowires is small and the recombination processes are fast enough in comparison with period of heating[28,29].

It should be noted that the effects of thermal interface resistance could be neglected because the ratio of the nano-Si sample thickness relative to its thermal conductivity (~$10^{-5}$ m$^2$ K/W) was much higher for all series of samples than the typical thermal interface resistance value (~$10^{-7}$ m$^2$ K/W) [30–32].

Since, the voltage $U$ detected from piezoelectric transducer electrodes is proportional to the thermoelastic stresses inside the sample[33], photoacoustic response ($U(\omega)$) depends on the temperature rise induced by the absorption of radiation. The correlation of the experimental and theoretical values of such temperature rise allows to evaluate the thermal diffusivity of the material D through inverse modelling.

### 3. Results and discussion

The typical amplitude-frequency characteristics of the photoacoustic response for the cases with and without water covering layer are shown in the Fig. 4. Photoacoustic characteristics for porous silicon samples with different porosities are presented in Fig. 4a, b while Fig. 4c, d stand for silicon nanowire arrays. In both materials the amplitude dependencies with frequency of photoacoustic signals change with porosity or thickness of layers. As we can see, there is a change of the curve slope on the amplitude-frequency variation for both series of samples. Such behavior is related to the different behavior of thermal and mechanical properties of the studied samples above the silicon substrate. The results of numerical simulations using the theoretical model described in the previous section are plotted with continuous lines (3) in Fig. 4. Simulation results are in good agreement with the experimental data. The difference between photoacoustic formation mechanisms in porous silicon and silicon nanowires arrays analyzed in detail in [34].

Additional calorimetric measurements are required for evaluation of the heat capacity $C_s$ and the thermal conductivity $\kappa_s$ of nanostructured systems. For this purpose, the surface of the sample was covered with a liquid layer. As a result, the optical deposited energy is redistributed at the surface of the structure. The heat fluxes at the interface "nanostructured solid/fluid" should be equal in accordance with the Fourier law:

$$\kappa_l \frac{\partial T}{\partial z}\bigg|_{z=0-0} = \kappa_s \frac{\partial T}{\partial z}\bigg|_{z=0+0}$$

where $\kappa_l$, $\kappa_s$ are the thermal conductivities of the liquid and the sample.

In the case of the dry sample all thermal energy is concentrated inside porous silicon or silicon nanowires layer. In the case with liquid film energy redistributed between liquid and solid in contact. The ratio of thermal energy at the interface is determined by the correlation of the thermal conductivities of the contact media. Furthermore, the amount of thermal energy that penetrates inside the sample depends on the thermal conductivity of the studied material. As it was mentioned above, the parameters of the photoacoustic response depend on the source of thermoelastic stresses, which will be modified due to the heat outflow into the liquid layer.

The typical amplitude-frequency characteristics of the photoacoustic response in the cases of absence (1) and presence (2) of the surface liquid layer are shown in Fig. 4. In the presence of such layer, the amplitude of the photoacoustic response systematically decreases (2) for both series of the studied samples and the reduction is more important in increasing frequency.

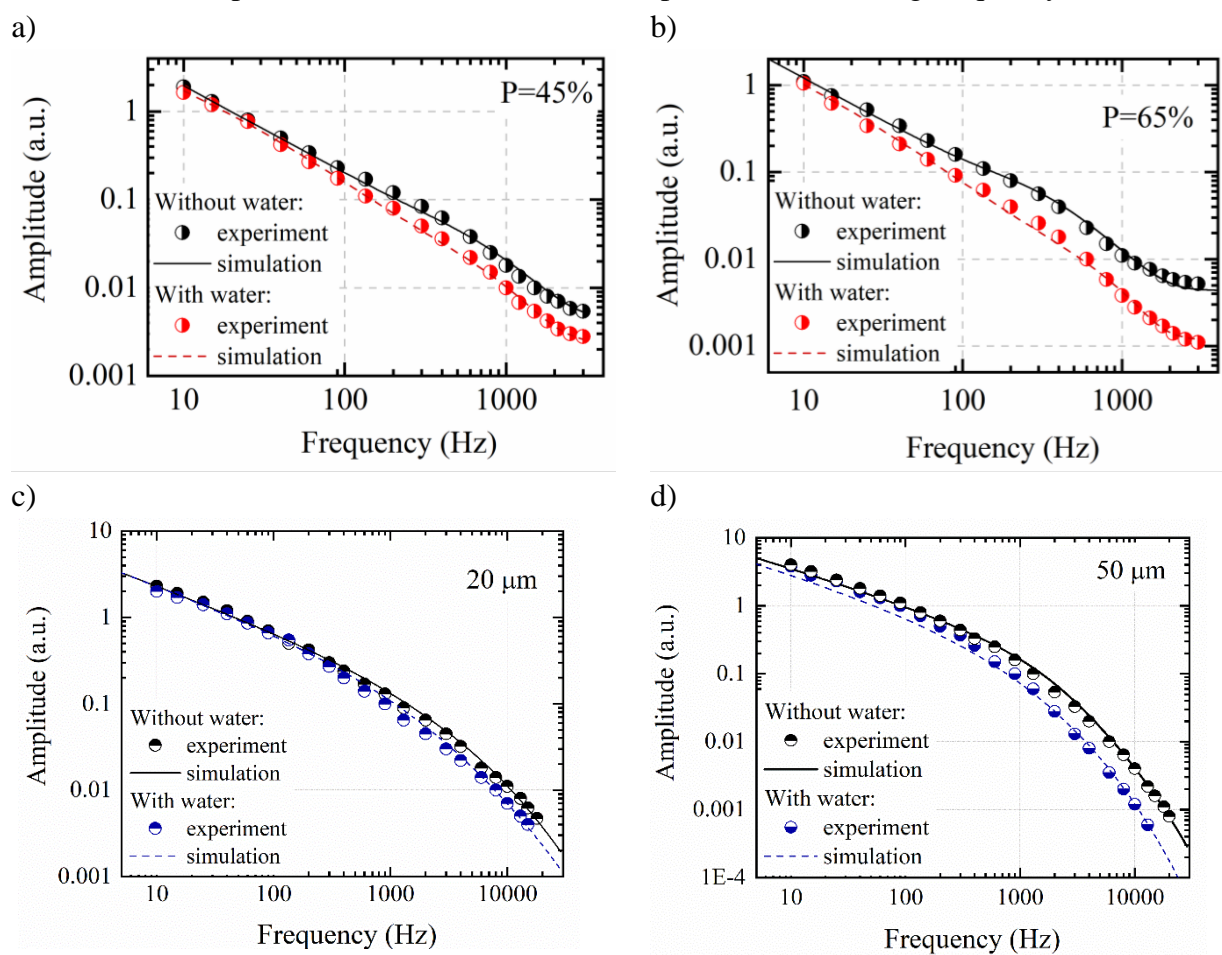

Fig. 4 The amplitude-frequency characteristics of the photoacoustic response of the studied system with porous silicon layer (a, b) and silicon nanowires arrays (c, d) for the cases of absence and presence covering liquid layer.

It should be noted that such decrease of the amplitude depends on the thickness of the sample. In the case of respectively thin layer of PS or SiNWs, the heat perturbation effectively penetrates in the substrate with a high value of thermal conductivity ($\kappa_{Si} \gg \kappa_s, \kappa_{Si} \gg \kappa_l$). Thus, influence of interface between nanostructured solid and fluid is not so crucial, especially at low frequency range. With increasing frequency, the length of the thermal wave decreases ($\lambda = \sqrt{a/2\omega}$)[35], and heat perturbation in this case is localized in PS or SiNWs.

Therefore, as one can see in Fig. 4, the amplitudes of photoacoustic signals in the cases of absence and occurrence of the water layer are approximately the same for frequencies $< a l_s^2$. With increasing frequency, when the length of the thermal wave becomes commensurate to the sample thickness, the thermal energy is more efficiently released from the silicon nanostructures inside liquid.

Dependencies of the ratio of frequency amplitudes "ξ" (see Fig. 5) for the wet sample and the pristine one are more informative. In this figure, one can see that the magnitude of the PA signal decrease is due to the presence of the covering liquid film. The decrease of the magnitude has a well-pronounced minimum. The position of the minimum depends on the thickness of the nanostructured layer as well as the thermal physical properties. Figure 3 shows quite good agreement between simulations and experimental results. It is also necessary to mention at this point, that experimental points for a thinner sample of SiNWs arrays are closer to the simulated curve corresponding to porosity of 50 %, while the points for a thickest sample are closer to the simulated curve corresponding to porosity of 60 %. The latter corresponds well with the above-mentioned assumption that during fabrication of SiNWs porosity slightly increases.

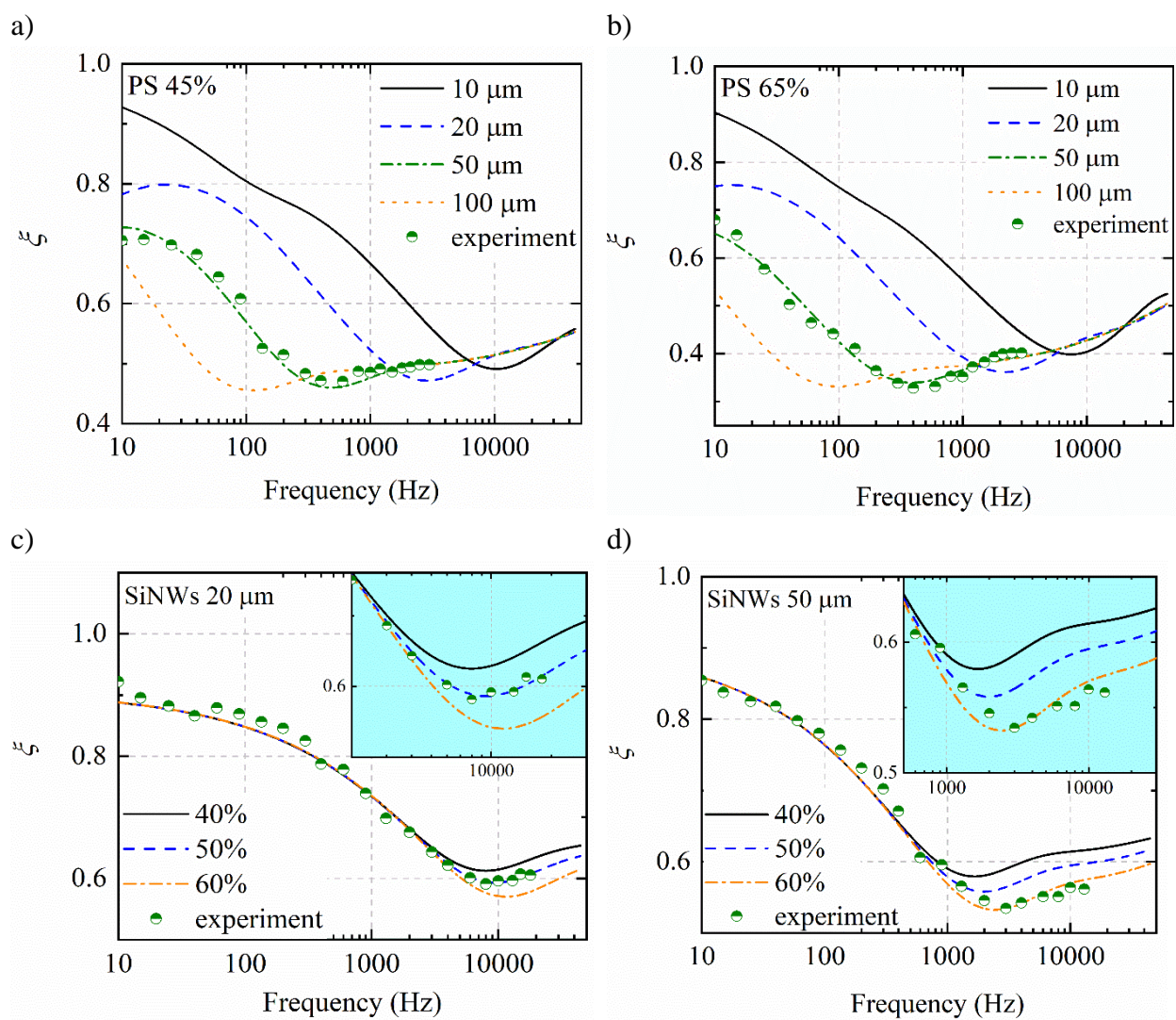

Fig. 5. Experimental (by dots) and simulated (by lines) frequency dependencies of the amplitude ratio measured for the sampled covered by liquid and the initial one. The simulated curves are presented for different thicknesses of the porous silicon with porosities 45 % (a) and 65 % (b), and for different porosities of silicon nanowires arrays with thicknesses equal to 20 μm (c) and 50 μm (d).

Correlations of two cases, defined by the presence and the absence of covering liquid layer gives us the possibility for simultaneous evaluation of the values of the heat capacity $C_s$ and of the thermal conductivity $\kappa_s$. Such evaluation was performed by the fitting of experimental amplitude-frequency characteristics with simulated ones. The extracted and literature values are shown in Table 1 for porous silicon samples and in Table 2 for silicon nanowires arrays.

Additionally, obtained results presented in Fig. 6.

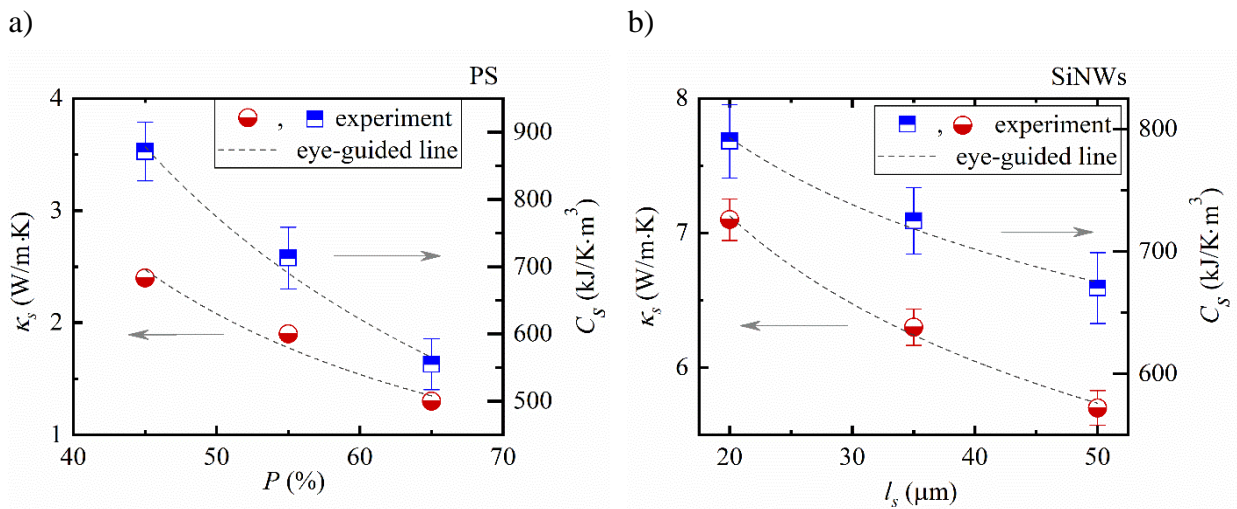

Fig. 6 Thermophysical parameters of the porous silicon samples with different porosity (a) and the silicon nanowires arrays with different thickness (b).

Thermal conductivity of the porous silicon samples decreases with porosity according to the following expression: $\kappa_s = (1-P)^3 \kappa_{Si}$.[36] As one can see, the heat capacity also decreases with increasing porosity of the layer. The values of heat capacities obtained using the photoacoustic approach agree with those that can be estimated from the approximation of the effective medium: $\rho_s C_s = (1-P)\rho_{Si} C_{Si}$.

Table 1 - Thermophysical parameters of the porous silicon samples with different porosity

| № | $P$, % | $C_s$, kJ/K·m³ | Experiment $\kappa_s$, W/m·K | Literature data $\kappa$, W/m·K |
|---|---|---|---|---|
| 1 | 45 ± 5 | 871 ± 79 | 2.4 ± 0.3 | 2.8[37] |
| 2 | 55 ± 7 | 713 ± 110 | 1.9 ± 0.3 | 1.4[37] |
| 3 | 65 ± 6 | 555 ± 95 | 1.3 ± 0.3 | 1.0[37] |

Table 2 - Thermophysical parameters of the silicon nanowires arrays with different thickness

| № | $l_s$, μm | $C_s$, kJ/K·m³ | Experiment $\kappa_s$, W/m·K | Literature data $\kappa$, W/m·K |
|---|---|---|---|---|
| 1 | 20 | 792 ± 63 | 7.1 ± 0.4 | 8[38]; 5.8[39] |
| 2 | 35 | 745 ± 76 | 6.3 ± 0.4 | 4.2[39] |
| 3 | 50 | 670 ± 79 | 5.7 ± 0.4 | 5.4[38] |

The heat capacity of the nanowires arrays decreases with increasing thickness of layer. However, since the average dimensions of the nanowires diameter (~ 100 nm), it can be assumed that the value of the specific heat is the same as the specific heat of crystalline silicon[40]. Consequently, the change in the heat capacity can be related with variation of the porosity of the etched nanowires as it was observed on the Fig. 5.

As can be seen from Table 2, the value of thermal conductivity decreases with increasing SiNWs layer thickness. Such decrease can be explained by increasing the structure heterogeneity, broken nanowires and porosity arising during the etching process [41]. This consideration is corelated with the previous assumption regarding the reducing of heat capacity with increasing of the thickness of the layer.

## 4. Conclusions

An experimental method based on the heat fluxes redistribution on the surface of the studied structures is proposed in order to appraise simultaneously material heat capacity and thermal conductivity. This method is based on the deposition of covering liquid layer on the sample surface, which acts like a heat sink and leads to redistribution of heat fluxes. In this framework, the photoacoustic technique with piezoelectric recording has been chosen to achieve nondestructive contactless thermal properties diagnostics. The features of the amplitude-frequency characteristics of the photoacoustic response for silicon-based nanostructured materials with different morphology are analyzed with and without covering liquid film.

Firstly, the approach was tested on the samples of porous silicon with different porosities. The porous silicon was chosen because of the possibility to precisely control the thickness of the sample and its porosity during the fabrication processes. The experimental results are in good agreement with the proposed theoretical model of the photoacoustic response formation in such structures.

Then, the methodology was applied for the silicon nanowire arrays fabricated by metal assisted chemical etching. It was found that thermal conductivity and the heat capacity of the samples decreased with increasing the etching time. Considering the average sizes of nanowires diameter, one can assume that the specific heat capacity of SiNWs arrays is the same as the one of crystalline silicon. Therefore, such behavior of the heat capacity could be associated with the porosity rise of the studied materials as it was observed on the amplitude frequency dependencies. The decrease of the thermal conductivity could be also explained by the increase of the porosity and the growth of the structure heterogeneities occurred during the silicon substrate etching.

**Supplementary Materials**

Supplementary materials containing SEM images of all samples used in the work.

**Acknowledgment**

The publication contains the results obtained in the frames of research work "Features of photothermal and photoacoustic processes in low-dimensional silicon-based semiconductor systems" (Ministry of Education and Science of Ukraine, State Registration Number 0118U000242) and project HotLine (Agence nationale de la recherche, France, project number ANR-19-CE09-0003). This work has been partially funded by the CNRS Energy unit (PEPS Cellule ENERGIE 2019) through the project Im-HESurNaASA. Mykola Isaiev and David Lacroix want also to acknowledge the partial financial support of the scientific pole EMPP of University of Lorraine.

**Data Availability**

The data that support the findings of this study are available from the corresponding author upon reasonable request.

**References:**

[1] Y. V Ryabchikov, V. Lysenko, and T. Nychyporuk, J. Phys. Chem. C **118**, 12515 (2014).

[2] A. Al-Kattan, Y. V Ryabchikov, T. Baati, V. Chirvony, J.F. Sanchez-Royo, M. Sentis, D. Braguer, V.Y. Timoshenko, M.-A. Esteve, and A. V Kabashin, J. Mater. Chem. B **4**, 7852 (2016).

[3] O. Olikh and K. Voytenko, Ultrasonics **66**, (2016).

[4] C.F. Ramirez-Gutierrez, J.D. Castaño-Yepes, and M.E. Rodriguez-Garcia, Optik (Stuttg). **173**, 271 (2018).

[5] C.F. Ramirez-Gutierrez, A. Medina-Herrera, L. Tirado-Mejía, L.F. Zubieta-Otero, O. Auciello, and M.E. Rodriguez-Garcia, J. Lumin. **201**, 11 (2018).

[6] D.G. Cahill, K. Goodson, and A. Majumdar, J. Heat Transfer **124**, 223 (2002).

[7] S. Gomès, A. Assy, and P.O. Chapuis, Phys. Status Solidi Appl. Mater. Sci. **212**, 477 (2015).

[8] D.G. Cahill, Rev. Sci. Instrum. **61**, 802 (1990).

[9] L. Lu, W. Yi, and D. Zhang, Rev. Sci. Instrum. **72**, 2996 (2001).

[10] F. Volklein, Thin Solid Films **188**, 27 (1990).

[11] M. Okuda and S. Ohkubo, Thin Solid Films **213**, 176 (1992).

[12] M. Gustavsson, E. Karawacki, and S.E. Gustafsson, Rev. Sci. Instrum. **65**, 3856 (1994).

[13] T.M. Tritt, *Thermal Conductivity Theory, Properties, and Applications*, Kluwer Aca (Plenum Publishers, New York 233 Spring Street, New York, New York 10013, 2004).

[14] C. Chiritescu, Science (80-. ). **315**, 351 (2007).

[15] G. Chen, J. Nanoparticle Res. **2**, 199 (2000).

[16] K. Termentzidis, *Nanostructured Semiconductors: Amorphization and Thermal Properties* (Pan Stanford, 2017).

[17] P. Lishchuk, D. Andrusenko, M. Isaiev, V. Lysenko, and R. Burbelo, Int. J. Thermophys. **36**, 2428 (2015).

[18] R. Burbelo, D. Andrusenko, M. Isaiev, and A. Kuzmich, Arch. Metall. Mater. **56**, 1157 (2011).

[19] J.A. Balderas-López, Rev. Sci. Instrum. **77**, (2006).

[20] K.G. Biswas, T.D. Sands, B.A. Cola, and X. Xu, Appl. Phys. Lett. **94**, 6 (2009).

[21] B. Abad, M. Rull-Bravo, S.L. Hodson, X. Xu, and M. Martin-Gonzalez, Electrochim. Acta **169**, 37 (2015).

[22] Z. Šoškić, S. Ćirić-Kostić, and S. Galović, Int. J. Therm. Sci. **109**, 217 (2016).

[23] S. Alekseev, D. Andrusenko, R. Burbelo, M. Isaiev, and a Kuzmich, J. Phys. Conf. Ser. **278**, 012003 (2011).

[24] M. Isaiev, K. Voitenko, V. Doroshchuk, D. Andrusenko, A. Kuzmich, A. Skryshevskii, V.

Lysenko, and R. Burbelo, in *Phys. Procedia* (2015).

[25] D. Andrusenko, M. Isaiev, A. Kuzmich, V. Lysenko, and R. Burbelo, Nanoscale Res. Lett. **7**, 411 (2012).

[26] M. Isaiev, D. Andrusenko, A. Tytarenko, A. Kuzmich, V. Lysenko, and R. Burbelo, Int. J. Thermophys. **35**, 2341 (2014).

[27] K. Voitenko, D. Andrusenko, A. Pastushenko, M. Isaiev, A.G. Kuzmich, and R.M. Burbelo, J. Nano- Electron. Phys. **9**, (2017).

[28] D.K. Markushev, D.D. Markushev, S. Aleksić, D.S. Pantić, S. Galović, D.M. Todorović, and J. Ordonez-Miranda, J. Appl. Phys. **126**, (2019).

[29] D.D. Markushev, J. Ordonez-Miranda, M.D. Rabasović, S. Galović, D.M. Todorović, and S.E. Bialkowski, J. Appl. Phys. **117**, (2015).

[30] E.T. Swartz and R.O. Pohl, Rev. Mod. Phys. **61**, 605 (1989).

[31] A. V Chernatynskiy and D.H. Hurley, J. Appl. Phys. **115**, (2014).

[32] H. Wang, Y. Xu, M. Shimono, Y. Tanaka, and M. Yamazaki, Mater. Trans. **48**, 2349 (2007).

[33] K. Dubyk, L. Chepela, P. Lishchuk, A. Belarouci, D. Lacroix, and M. Isaiev, Appl. Phys. Lett. **115**, 021902 (2019).

[34] K. Dubyk, A. Pastushenko, T. Nychyporuk, R. Burbelo, M. Isaiev, and V. Lysenko, J. Phys. Chem. Solids **126**, 267 (2019).

[35] H.S. Carslaw and J.C. Jaeger, *Conduction of Heat in Solids*, 2nd ed. (Clarendor Press, Oxford, 1959).

[36] P.J. Newby, in *Porous Silicon From Form. to Appl. Form. Prop. Vol. One*, edited by G. Korotcenkov (CRC Press, 2016), pp. 237–250.

[37] V. Lysenko, L. Boarino, M. Bertola, B. Remaki, A. Dittmar, G. Amato, and D. Barbier, **30**, 1141 (1999).

[38] A.I. Hochbaum, R. Chen, R.D. Delgado, W. Liang, E.C. Garnett, M. Najarian, A. Majumdar, and P. Yang, Nature **451**, 163 (2008).

[39] M. Isaiev, O. Didukh, T. Nychyporuk, V. Timoshenko, and V. Lysenko, Appl. Phys. Lett. **110**, (2017).

[40] A.A.R. A.I. Gusev, in *Nanocrystalline Mater.*, 1st editio (Cambridge International Science Publishing Ltd;, Cambridge, 2004), pp. 177–190.

[41] S.P. Rodichkina, L.A. Osminkina, M. Isaiev, A. V. Pavlikov, A. V. Zoteev, V.A. Georgobiani, K.A. Gonchar, A.N. Vasiliev, and V.Y. Timoshenko, Appl. Phys. B Lasers Opt. **121**, 337 (2015).